\documentclass{aastex62}

\usepackage{graphicx}
\usepackage{lineno}

\graphicspath{{./}{figures/}}

\shorttitle{The reddening map of Liller 1}
\shortauthors{Pallanca et al.}

\begin{document}

\title{High-resolution extinction map in the direction of the strongly
  obscured bulge fossil fragment Liller 1}\footnote{Based on
observations collected with the NASA/ESA HST (Prop. GO 15231),
obtained at the Space Telescope Science Institute, which is operated
by AURA, Inc., under NASA contract NAS5-26555.  Based on observations
(Prop. GS-2013-Q-23) obtained at the Gemini Observatory, which is
operated by the Association of Universities for Research in Astronomy,
Inc., under a cooperative agreement with the NSF on behalf of the
Gemini partnership: the National Science Foundation (United States),
the National Research Council (Canada), Agencia Nacional de
Investigaci\'on y Desarrollo (Chile), the Australian Research Council
(Australia), Minist\'erio da Ci\^encia, Tecnologia e Inova\c{c}\~ao
(Brazil), and Ministerio de Ciencia, Tecnolog\'ia e Innovaci\'on
Productiva (Argentina).  Based on observations gathered with the
ESO-VISTA telescope (program ID 179.B-2002). }

\correspondingauthor{Cristina Pallanca}
\email{cristina.pallanca3@unibo.it}
\author[0000-0002-7104-2107]{Cristina Pallanca} \affil{Dipartimento di
  Fisica e Astronomia, Universit\`a di Bologna, Via Gobetti 93/2,
  Bologna I-40129, Italy} \affil{Istituto Nazionale di Astrofisica
  (INAF), Osservatorio di Astrofisica e Scienza dello Spazio di
  Bologna, Via Gobetti 93/3, Bologna I-40129, Italy}
\author[0000-0002-2165-8528]{Francesco R. Ferraro} \affil{Dipartimento
  di Fisica e Astronomia, Universit\`a di Bologna, Via Gobetti 93/2,
  Bologna I-40129, Italy} \affil{Istituto Nazionale di Astrofisica
  (INAF), Osservatorio di Astrofisica e Scienza dello Spazio di
  Bologna, Via Gobetti 93/3, Bologna I-40129, Italy}
\author[0000-0001-5613-4938]{Barbara Lanzoni} \affil{Dipartimento di
  Fisica e Astronomia, Universit\`a di Bologna, Via Gobetti 93/2,
  Bologna I-40129, Italy} \affil{Istituto Nazionale di Astrofisica
  (INAF), Osservatorio di Astrofisica e Scienza dello Spazio di
  Bologna, Via Gobetti 93/3, Bologna I-40129, Italy}
\author[0000-0000-0000-0000]{Chiara crociati} \affil{Dipartimento di
  Fisica e Astronomia, Universit\`a di Bologna, Via Gobetti 93/2,
  Bologna I-40129, Italy}\affil{Istituto Nazionale di Astrofisica
  (INAF), Osservatorio di Astrofisica e Scienza dello Spazio di
  Bologna, Via Gobetti 93/3, Bologna I-40129, Italy}
\author[0000-0003-4746-6003]{Sara Saracino} \affil{ Astrophysics
  Research Institute, Liverpool John Moores University, 146 Brownlow
  Hill, Liverpool L3 5RF, UK} \author[0000-0003-4237-4601]{Emanuele
  Dalessandro} \affil{Istituto Nazionale di Astrofisica (INAF),
  Osservatorio di Astrofisica e Scienza dello Spazio di Bologna, Via
  Gobetti 93/3, Bologna I-40129, Italy}
\author[0000-0002-6040-5849]{Livia Origlia} \affil{Istituto Nazionale
  di Astrofisica (INAF), Osservatorio di Astrofisica e Scienza dello
  Spazio di Bologna, Via Gobetti 93/3, Bologna I-40129, Italy}
\author[0000-0003-0427-8387]{Michael R. Rich} \affil{Department of
  Physics and Astronomy, UCLA, PAB 430 Portola Plaza, box 951547, LA
  CA 90095-1547, USA }
\author[0000-0002-6092-7145]{Elena Valenti}
\affil{European Southern Observatory, Karl-Schwarzschild-Stra\ss e 2,
  D-85748 Garching bei M\"unchen, Germany}\affil{Excellence Cluster
  ORIGINS, Boltzmann-Stra\ss e 2, D-85748 Garching bei M\"unchen,
  Germany} \author[0000-0002-3900-8208]{Douglas Geisler}
\affil{Departamento de Astronom\'ia, Universidad de Concepci\'on,
  Casilla 160 C, Concepci\'on, Chile}\affil{Instituto de
  Investigaci\'on Multidisciplinario en Ciencia y Tecnolog\'ia,
  Universidad de LaSerena. Avenida Ra\'ul Bitr\'an S/N, La Serena,
  Chile}\affil{Departamento de F\'isica y Astronom\'ia, Facultad de
  Ciencias, Universidad de La Serena. Av.Juan Cisternas 1200, La
  Serena, Chile} \author[0000-0000-0000-0000]{Francesco Mauro}
\affil{Instituto de Astronom\'ia, Universidad Cat\'olica del Norte,
  Av. Angamos 0610, Antofagasta, Chile}
\author[0000-0001-6205-1493]{Sandro Villanova} \affil{Departamento de
  Astronom\'ia, Universidad de Concepci\'on, Casilla 160 C,
  Concepci\'on, Chile} \author[0000-0000-0000-0000]{Christian Moni
  Bidin} \affil{Instituto de Astronom\'ia, Universidad Cat\'olica del
  Norte, Av. Angamos 0610, Antofagasta, Chile}
\author[0000-0002-3865-9906]{Giacomo Beccari} \affil{European Southern
  Observatory, Karl-Schwarzschild-Stra\ss e 2, D-85748 Garching bei
  M\"unchen, Germany}

\begin{abstract}
We used optical images acquired with the Wide Field Camera of the
Advanced Camera for Surveys onboard the {\it Hubble Space Telescope}
and near-infrared data from GeMS/GSAOI to construct a high-resolution
extinction map in the direction of the bulge stellar system Liller
1.  In spite of its appearance of a globular cluster, Liller 1 has
  been recently found to harbor two stellar populations with
  remarkably different ages, and it is the second complex stellar
  system with similar properties (after Terzan5) discovered in the
  bulge, thus defining a new class of objects: the Bulge Fossil
Fragments. Because of its location in the inner bulge of the Milky
Way, very close to the Galactic plane, Liller 1 is strongly affected
by large and variable extinction.
 The simultaneous study of both the optical and the near-infrared
  color-magnitude diagrams revealed that the extinction coefficient
  R$_V$ in the direction of Liller 1 has a much smaller value than
  commonly assumed for diffuse interstellar medium (R$_V=2.5$, instead
  of 3.1), in agreement with previous findings along different light
  paths to the Galactic bulge.  The derived differential reddening
map has a spatial resolution ranging from $1 \arcsec$ to $3 \arcsec$
over a field of view of about $90 \arcsec \times 90 \arcsec$.  We found
that the absorption clouds show patchy sub-structures with extinction
variations as large as $\delta {\rm E}(B-V)\sim0.9$
mag.  \end{abstract}


\section{INTRODUCTION}
\label{intro}
Liller 1 is a stellar system historically cataloged as a globular
cluster (GC) located in the innermost region of the Galaxy (at only
0.8 kpc from the Milky Way center). It lies very close to the Galactic
plane ($l=354.84, b=-0.16$; \citealt{harris}), in a region that is
affected by large and spatially variable foreground extinction. This
is the main reason hindering detailed photometric investigations of
its stellar population.  One of the first works about the
  extinction toward Liller 1 dates back to the eighties, when
  \citet{armandroff88} estimated a color excess E$(B-V)=2.7$. Then,
  \citet{frogel95} found E$(J-K)=1.7$, corresponding to extinction
  coefficients $A_K=0.87$, $A_V=9.5$ and $E(B-V)=3.0$. In the same
  work the authors analyzed an offset field and found a larger
  extinction coefficient,
  thus suggesting the presence of differential reddening.
  Later works, both in the optical \citep{ortolani96} and mainly in
  the near-infrared (NIR; \citealp{ortolani01, barbuy98, davidge00,
    valenti10}), found values of E$(B-V)$ ranging from 3 up to 3.13,
  and also highlighted the presence of differential absorption along
  the line of sight. \citet{gonzalez1,gonzalez2} used data from the
  VISTA Variables in the V\'ia L\'actea (VVV) Survey \citep{minniti10}
  and derived the mean $(J -K)$ color of red clump giants in 1835
  sub-fields in the bulge region (also including the area covered by
  Liller 1); then, through the comparison with the color of similar
  stars in the Baade's window, they derived a reddening map with a
  maximum resolution of $2\arcmin$.
  According to the online simulator\footnote{\it
  http://mill.astro.puc.cl/BEAM/calculator.php} based on this work, a
  color excess E$(J-K)=1.42\pm 0.16$, corresponding to E$(B-V)=2.70$
  is found in the direction of Liller 1.  More recently, by using the
  multi-conjugate adaptive-optics system GSAOI at the GEMINI South
  Telescope, \citet{saracino15} obtained the first insight into the
  cluster population down to its main sequence (MS) turnoff level.
  They built a reddening map of the central region of Liller 1 with a
  a spatial resolution of $18\arcsec \times 18\arcsec$, finding a mean
  color excess E$(B-V)=3.3\pm0.2$ and a maximum variation $\delta {\rm
    E}(B-V)=0.34$ within the surveyed field of view (FOV).
The dataset was also used to accurately determine the structural and
physical parameters (scale radii, concentration parameter, central
mass density and total mass) of Liller 1, confirming that this is a
massive stellar system (of 1-2 $\times 10^6 M_\odot$) with a very
large collision rate (the second-highest after Terzan 5).

Recently, our group performed the very first {\it Hubble Space
  Telescope} (HST) observations of Liller 1 in the optical band,
securing a set of deep Advanced Camera for Surveys (ACS) images of its
central region. By combining these data with the GEMINI NIR
observations, \citet[][hereafter F21]{ferraro21} discovered the
presence of two distinct stellar sub-populations with remarkably
different ages: one as old as $\sim 12$ Gyr, thus named ``Old
Population (OP)'', and another one much younger, with an age down to
only 1 Gyr, that we thus call
``young population (YP)''.  This discovery unquestionably demonstrates
that Liller 1 is not a genuine GC. It is the second stellar system in
the bulge \citep[after Terzan 5;][]{ferraro09} that appears as a GC,
but harbors, instead, a young sub-population of stars. The oldest
populations in Liller 1 and in Terzan 5 are impressively similar, in
agreement with the typical age of bulge GCs, thus indicating that
these systems formed at the same cosmic epoch, from gas clouds with
comparable chemistry. In turn, the chemical abundance patterns
measured in Terzan 5 are remarkably similar to those observed in the
bulge field stars \citep{ferraro16a, lanzoni10, origlia11, origlia13,
  origlia19, massari15}. These observational facts allowed F21 to
define a new class of stellar systems, the ``Bulge Fossil Fragments
(BFFs)'': these are stellar aggregates with the appearence of massive
GCs orbiting the Galactic bulge, formed at the epoch of the Galaxy
assembling, and harbouring in addition to the old population, a
younger (1-3 Gyr old) component. These stellar systems could, in fact,
be interpreted as the surviving relics of much more massive primordial
structures that generated the Galactic bulge (similar to the giant
clumps observed in the star forming regions of high-redshift galaxies;
\citealt{immeli04,carollo07,elmegreen08,genzel+11,tacchella+15,behrendt+16}).

Within this exciting scenario, we are now coordinating a project aimed
at reconstructing the origin and the star formation history of the
latest discovered BFF. Thus, while future specific papers will be
devoted to reconstruct the star formation history of Liller 1
(Emanuele Dalessandro et al., in preparation) and the chemical
abundance patterns of its sub-populations (Livia Origlia et al., in
preparation),  here we  determine the first high
  resolution differential reddening map in the direction of the
  system.  This map was used by F21 to construct the de-reddened
  color-magnitude diagram (CMD) that revealed the presence of the
  young population in Liller 1 and to study its properties.  The paper
  is organized as follows. In Section \ref{obs} we describe the
  dataset used in this work and the main steps of the photometric
  analysis. In Section \ref{redd} we present the determination of 
  the differential reddening map of Liller 1.  In Section
  \ref{discuss} we discuss the main results of the paper and summarize
  the conclusions.

\section{OBSERVATIONS AND DATA ANALYSIS}
\label{obs}
\subsection{Datasets}
The photometric dataset used for this work consists in a combination
of observations obtained at optical and NIR wavelengths.  By using the
Wide Field Channel (WFC) of the ACS on board the HST, we secured the
first set of high-resolution optical images ever acquired for Liller 1
(GO 15231, PI: Ferraro). The observations have been performed in the
filters F606W ($V$) and F814W ($I$), and they consist in a set of 6
deep images per filter (see Table \ref{Tab:dataset}). To study Liller 1 at NIR wavelengths we used
a dataset of high-resolution $J$ and $Ks$ (hereafter $K$) images
acquired with the multiconjugate adaptive-optics assisted camera
GeMS/GSAOI at the GEMINI South Telescope (GS-2-13-Q-23, PI:
Geisler). These images  were already used to obtain a first insight
into the stellar population in Liller1 by \citet{saracino15}.  
All the images are dithered by few pixels, in order
to avoid spurious effects due to bad pixels. In addition, the NIR
datasets were secured following a dither pattern large enough to fill
the inter-chip gaps, thus allowing a complete coverage of the
innermost $\sim 100\arcsec\times 100\arcsec$ region of the system.

\begin{table}
\begin{center}
\caption{Summary of the used datasets}
\begin{tabular}{l|l|c|c|c|l}
\hline
  Instrument & Program ID & PI & Date & Filter & ~~~~~~~~~~$\rm N_{exp} \times \rm t_{exp}$\\
                    &               &     & [yyyy/mm/dd] &  & \\             
\hline
 HST/ACS-WFC & GO 15231 & Ferraro &  2019/08/17 & F606W & $2 \times 1361 s$,~ $2 \times 1360 s$,~ $2 \times 1260 s$\\ 
             &          &         &             & F814W & $2 \times 855 s$,~ $1 \times 854 s$,~ $1 \times 837 s$,~ $2 \times 836 s$\\ 

\hline
\end{tabular}
\end{center}
\label{Tab:dataset}
\end{table}

\subsection{Photometric analysis, astrometry and magnitude calibration}
The data reduction of the ACS-WFC images has been performed on the
CTE-corrected (\texttt{flc}) images after a correction for
Pixel-Area-Map.  The photometric analysis was performed via the
point-spread function (PSF) fitting method, by using DAOPHOT
\citep{daophot} and following the ``standard'' approach used in
previous works \citep[e.g.,][]{pallanca17, dalessandro18,
  cadelanoM3}. Spatially variable PSF models were derived for each
image by using some dozens of bright and nearly isolated stars. The
PSF was modelled adopting an analytical \texttt{MOFFAT} function plus
a spatially variable look-up table and it has been applied to all the
identified sources with flux peaks at least 3 $\sigma$ above the local
background. We then built a master-list containing all the
stellar-like sources detected in more than 3 images in at least one
filter.
In order to improve the level of completeness, we performed
an accurate inspection of the residual images, 
in order to search for additional stellar sources missed during the first
iteration of the analysis because of the serious crowding conditions.
 This procedure allowed us to
recover a few dozens of stars, which were added to the master-list.
 
We then forced a fit at the position of each star belonging to the
master-list, in each frame, by using the \texttt{ALLFRAME} package
\citep[][]{daophot,allframe}. For each star, multiple magnitude
estimates from different exposures were homogenised by using DAOMATCH
and DAOMASTER, and their weighted mean and standard deviation were
finally adopted as the star magnitude and photometric error. The final
optical HST instrumental catalog contains all the stars measured in at
least 3 images in any of the two filters. For each star, it lists the
instrumental coordinates, the mean magnitude in each filter and two
quality parameters ({\it chi} and {\it sharpness}).\footnote{The {\it
  chi} parameter is the ratio between the observed and the expected
pixel-to-pixel scatter between the model and the profile image. The
{\it sharpness} parameter quantifies how much the source is spatially
different from the PSF model. In particular, positive values are
typical of more extended sources, such as galaxies and blends, while
negative values are expected in the case of cosmic rays and bad
pixels.}

\begin{figure*} 
\begin{center}
\includegraphics[width=190mm]{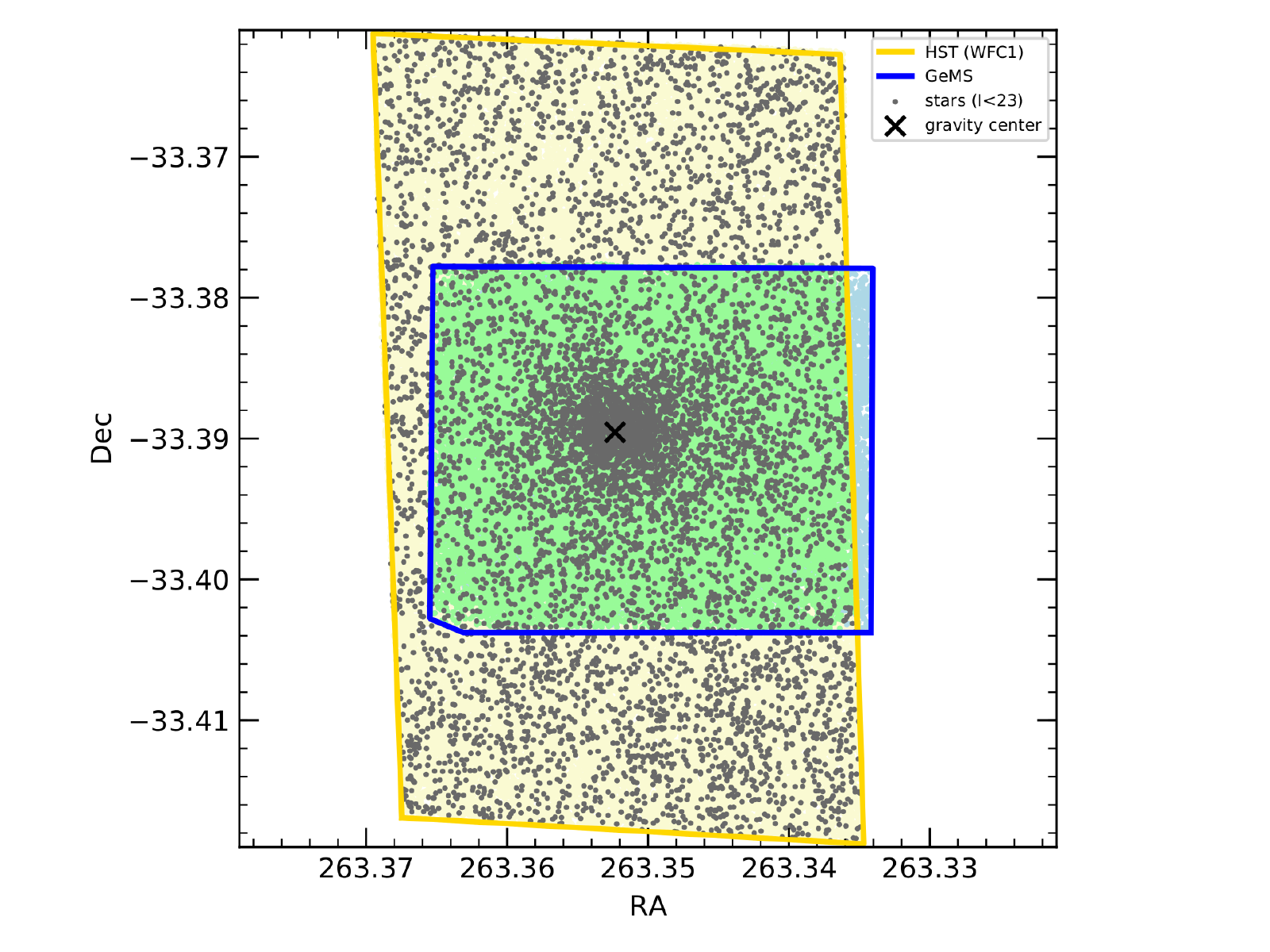}
\caption{Field of view sampled by the observations: in yellow all the
  stars within the rectangular area of $\sim 200\arcsec \times
  100\arcsec$ covered by chip1 of the ACS-WFC data-set, outlined in
  blue the objects detected within the $\sim 100\arcsec \times
  90\arcsec$ region corresponding to the GeMS/GSAOI observations, and
  in green the stars detected in the field of view in common between
  the two data-sets, for which we determined the differential
  reddening map. For reference, we also show the position of all the
  detected stars with $I<23$ (gray dots) and the gravity center
  (i.e. the photometric barycenter) of Liller 1 \citep[black
    cross;][]{saracino15}.  }
\label{Fig:fov}
\end{center}
\end{figure*}

The instrumental optical magnitudes were calibrated onto the VEGAMAG
photometric system by using the updated recipes and zero-points
available in the HST web-sites. The instrumental coordinates were
first corrected for geometric distortion and then reported to the
absolute coordinate system ($\alpha$, $\delta$) defined by the World
Coordinate System by using stars in common with the publicly available
Gaia DR2 catalog \citep{gaia16a,gaia16b}.  Although only disk field
stars have been found in common with the Gaia catalog (due to the very
high extinction toward this stellar system), the resulting
$1\sigma$ astrometric accuracy is $\sim 0.1\arcsec$. The reduction of
the ACS-WFC images has been limited to chip 1 because of the
negligible overlap between chip 2 and the FOV covered by the GeMS
observations (see Figure \ref{Fig:fov}).

At this point, the sample of stars detected in the optical and in the
NIR catalogs were compared. From a first analysis it was evident that
each catalog contained some objects missed in the other, mainly due to
the faintness of the source and/or its vicinity to very bright objects
with extreme (blue or red) colors. On the other hand, the analysis
also demonstrated that, because of the large extinction in the
direction of the system, the $V$-band exposures are significantly
shallower that those secured through the $I$ filter, and, similarly,
the $J$ images are much less deep than those obtained through the $K$
filter. This clearly indicated that the optimal solution for the
accurate study of the stellar populations in Liller 1 
including optical and NIR filters consisted in
combining the $I$ and the $K$ images.
We thus proceeded with a second-level analysis
starting from the independent reduction of 
the optical dataset and the 6 best-quality (per each filter) NIR images.
Two lists of sources have been obtained: one comprising all the stars
detected in the $I$ filter and one including the objects identified in
the $K$ band. The $I$-band list contained all the stars measured in at
least 3 frames, over the 6 available. The $K$-band list contained all
the stars measured in at least 3 over the 6  images
analysed.
We thus combined the two datasets, building a single master-list that
contains all the detected stars (some having magnitude measured in
both filters, others having only the $I$ or the $K$ measure). The
sources detected only in the $I$ images were then searched and
analysed also in the $K$ images, and vice versa, thus maximizing the
information provided by the two filters separately. The final
magnitudes of this combined analysis were homogenized to the
calibrated stand-alone HST and GeMS catalogs, in the Vegamag and 2MASS
systems, respectively.  The magnitudes of the brightest stars
($K<12.5$), which are saturated in the Gemini observations, have been
recovered (up to $K\sim10$) from the PSF photometry of VVV Survey data
performed by \citet{mauro13}. The final catalog reports the $I$ and
$K$ band magnitudes (together with the $V$ and $J$ ones, if measured)
for 43629 stars detected in both HST and GeMS images.

Figure \ref{Fig:fov} shows the FOV sampled by chip1 of the HST/ACS-WFC
(outlined in yellow) and by the GeMs observations (dashed blue
line). As apparent, the overlapping region, from which we can
construct the $(I, I-K)$ CMD and determine the differential reddening
map, corresponds to an area of about $\sim 90\arcsec \times 90\arcsec$
(green area), almost centered onto the gravity center (i.e. determined
as the photometric barycenter) of Liller 1 \citep{saracino15}.
Hereafter, all the discussion and plots are made considering only the
stars in this region.  Their CMD distribution is shown in three
different color and magnitude combinations in Figure
\ref{Fig:tripanelNC}: from left, to right, the optical $(I, V-I)$ CMD,
the ``hybrid'' $(I, I-K)$ CMD, and the NIR $(K, J-K)$ one.  As
expected, extinction along the line of sight limits the deepness of
the sample expecially in the optical diagram. The differential effect
of the reddening is particularly visible in the deformed morphology of
the red clump, which appears to be largely stretched along the
direction of the reddening vector (from top-left, to bottom-right),
especially in the optical and hybrid CMDs. This clearly shows that any
meaningful analysis of the stellar populations of Liller 1 requires an
appropriate modelling of and correction for the differential reddening
effects.

For almost all the stars in the catalog, we also measured the relative
proper motions by exploiting the time base-line of 6.3 yr separating
the available observations. This allowed us to distinguish the stars
belonging to Liller 1, from Galactic field interlopers (respectively,
black and orange dots in Figure \ref{Fig:tripanelNC}). The procedure
used for the proper motion measurement and the adopted selection
criteria for membership have been presented in F21.

\section{Differential reddening map}
\label{redd}
\subsection{Reddening laws}
\label{reddeningLaws}
As is well known, interstellar reddening is due to the absorption and
scattering of the radiation caused by dust clouds along the light path
to the observer. This phenomenon makes the flux of a given source
systematically fainter and the color systematically redder than their
true (emitted) values, with an efficiency that strongly depends on the
wavelength (increasing at shorter wavelengths).  The entity of
reddening is usually parametrised by the color excess E$(B-V)$,
defined as the difference between the observed color $(B-V)$ and the
intrinsic color $(B-V)_0$ that, for stars, is mainly fixed by the
temperature of the source.

The extinction law, i.e., the dependence of the absorption coefficient
on wavelength (A$_\lambda$) can be expressed as (see also the
formulations by \citealp{cardelli, odonnell94, fitzpatrick90,
  fitzpatrick99}):
\begin{equation}
{\rm A}_\lambda= {\rm R}_V \times c_{\lambda, {\rm R}_V} \times {\rm E}(B-V),
\end{equation}
 where $c_{\lambda, {\rm R}_V}=1$ is commonly adopted at the
  $V$-band wavelength ($c_{V, {\rm R}_V}=1$).  The parameter R$_V$ is
  usually set equal to 3.1, which is the standard value for diffuse
  interstellar medium  \citep{schultz75,sneden78}.
  This is indeed the value assumed
  in all previous works devoted to the estimate of reddening in the
  direction of Liller 1 (see the Introduction).  Particular attention,
  however, has to be given to the extinction toward the inner Galaxy
  \citep[][and references therein]{popowski00,
    nataf13,alonsoGarcia17,casagrande14,pallanca21param6440}, where the R$_V$ value seems to vary along
  different directions.  For example \citet{nataf13} suggested that
  R$_V=2.5$ is more appropriate to reproduce the bulge populations,
  while some other authors quote larger values \citep[e.g.,
    R$_V=3.2$;][]{bica16, kerber19}.

 Decreasing the value of R$_V$ makes the parameter $c_{\lambda,
    {\rm R}_V}$ (thus, the extinction law A$_\lambda/$A$_V$) more
  steeply dependent on wavelength. This is illustrated in the upper
  panel of Figure \ref{fig_RV}, where the two curves refer to the
  canonical R$_V$ value (R$_V=3.1$, blue line) and to R$_V=2.5$ (red
  line).  As shown in the central panel of the figure, the difference
  between $c_{\lambda,2.5}$ and $c_{\lambda,3.1}$ is quite small in
  the photometric bands considered in this work (marked by the dashed
  vertical lines).  However, this difference increases by one order of
  magnitude when these coefficients are multiplied by their respective
  value of R$_V$: this is shown in the bottom panel of the figure,
  plotting R$_{\lambda, {\rm R}_V} \equiv {\rm R}_V \times c_{\lambda,
    {\rm R}_V}$.  In turn, recalling that E$(B-V) = {\rm A}_\lambda
  /{\rm R}_{\lambda, {\rm R}_V}$, this implies that any given value of
  the absorption coefficient A$_\lambda$ corresponds to a
  significantly different color excess E$(B-V)$ if R$_V=2.5$ is
  assumed in place of the standard value R$_V=3.1$. Hence, the value
  of E$(B-V)$ and the distance modulus ($\mu_0$) necessary to
  superpose an isochrone onto an observed CMD change if R$_V=2.5$ is
  adopted instead of R$_V=3.1$.
  The bottom panel of Figure \ref{fig_RV} also shows that the
  difference between R$_{\lambda, 2.5}$ and R$_{\lambda, 3.1}$ is not
  constant, but significantly increases (in absolute value) for
  decreasing wavelength in the range relevant for the present study:
  it varies  between $\sim 0.1$ at $\lambda \simeq 21326.7$ \AA\ (the K
  band) and $\sim 0.6$ at $\lambda\simeq 5810.8$ \AA\ (the effective
  wavelength of the F606W filter; \citealt{lambdafilters1,lambdafilters2}).\footnote{ Following the
  prescriptions of \citet{cardelli} and \citet{odonnell94}, we find
  that, if R$_V=3.1$, the corresponding values in the four photometric
  bands here considered are: R$_{606}=2.89$, R$_{814}=1.90$,
  R$_J=0.89$, and R$_K=0.37$.  If R$_V=2.5$, then: R$_{606}=2.30$,
  R$_{814}=1.43$, R$_J=0.64$, and R$_K=0.27$. }  In turn, the
  increasing difference between R$_{\lambda, 2.5}$ and R$_{\lambda,
    3.1}$ for decreasing wavelength necessarily implies that two
  isochrones, the one shifted to the observed plane by adopting
  R$_V=3.1$, the other shifted under the assumption of R$_V=2.5$,
  cannot simultaneously reproduce the NIR and the optical CMDs of a
  stellar system.
  This is illustrated in Figure \ref{fig_isoCMDobs}, where the optical
  $(I, V-I$), hybrid $(I, I-K)$ and NIR $(K, J-K)$ CMDs of Liller 1
  are shown together with an
isochrone \citep{marigo17} computed for an age $t=12$ Gyr and a global
metallicity [M/H]$=-0.3$. The black dashed line shows the location in
the three CMDs of the isochrone obtained by assuming R$_V=3.1$,
E$(B-V)=3.30$ and $\mu_0=14.55$, which are the values needed to make
it reproduce the observed NIR CMD \citep{saracino15}. However, no
reasonable match of the observations can be simulatenously obtained in
the optical and hybrid CMDs. Conversely, by adopting R$_V=2.5$ (red
line) all the evolutionary sequences of the cluster in any wavelength
combination are well reproduced. This demonstrates that the canonical
reddening law is not appropriate in that direction of the sky, and a
much lower value of R$_V$ is needed toward Liller 1, in agreement with
what suggested in the pioneering work of \citep{frogel95} and what
previously found for other bulge directions
\citep[e.g.,][]{popowski00, nataf13, alonsoGarcia17,casagrande14,saha19,pallanca21param6440}.

The manifest inadequacy of R$_V=3.1$ in the direction of Liller 1 came
not into light in previous works because they were all based on NIR or
optical datasets separately: it is indeed the simultaneous analysis of
these CMDs that finally revealed the necessity of a much lower value
of R$_V$. Of course, however, a proper comparison between theoretical
isochrones and observations first requires to correct the CMDs for the
effects of differential reddening, which is responsible for the
distortions of the evolutionary sequences clearly visible in Figure
\ref{fig_isoCMDobs}.  
 
Our result (i.e., R$_V$=2.5) is perfectly in agreement with previous works aimed to estimate the reddening law in the Bulge direction \citep{nataf13,saha19}. For example, R$_V$=2.5 perfectly corresponds to the peak of the distribution shown in Figure 8 of \citet{nataf13}: indeed, as explicitly discussed in the conclusions of that paper, the mean value of the distribution, $A_I/E(V-I)\sim1.217$, corresponds to R$_V$=2.5. On the other hand, we found that the absorption coefficients derived by \citet{saha19} in the direction of the Baade's window and shown as black circles in their Figure 8 are well reproduced by the \citet{odonnell94} extinction law calculated for R$_V\sim2.0$, and the values obtained for R$_V$=2.5 place in the middle between these points and those corresponding to the standard R$_V$=3.1 (red circles in their figure). Note that a value as small as R$_V$=2.0, even if surprising, is still in agreement with the left-hand tail of the distribution shown in Figure 8 of \citet{nataf13}.  Hence, our result is perfectly in agreement with previous works and it brings further evidence that the commonly adopted R$_V$=3.1 is likely inappropriate toward the Bulge, while a precise estimate of the extinction law in each specific direction is required \citep[see also][]{pallanca21param6440}.

\subsection{The differential reddening map of Liller 1}
\label{diff_redd}
The interstellar extinction along the Galactic plane and in the
direction of the Galactic center can be highly spatially variable
(differential reddening), on scales that can be as small as a few
arcseconds. Since this spatial dimension is significantly smaller than
the apparent size of most Galactic stellar systems, the CMD of star
clusters affected by differential reddening shows a systematic
elongation of all the evolutionary sequences along the direction of
the reddening vector. In turn, this prevents the accurate
characterization of their photometric properties, with a non
negligible impact on the determination of fundamental parameters such
as, for instance, the distance and age \citep[e.g.,][]{bonatto13}.  To
model and correct for the effect of differential reddening in GCs, two
main approaches have been adopted in the literature.  Schematically:
(1) the ``cell by cell'' method, which consists in dividing the
observed FOV in a number of cells, build the CMD of the stars included
in each cell, and calculate the reddening value in each cell from the
color and magnitude displacements with respect to the bluest CMD
\citep{heitsch99, piotto99, vonbrau01, mcwilliam10, nataf10,
  gonzalez1, gonzalez2, massariTer5, bonatto13, saracino15}, and (2)
the ``star by star'' approach, which consists in estimating the
reddening of each star in the catalog from the relative displacement
between its ``local CMD'', i.e., the CMD built with the spatially
closest objects, and the cluster mean ridge line \citep{milone12,
  bellini13, saracino6569, pallanca6440, cadelano20}.   In the
  cited papers, different authors used stars in different evolutionary
  sequences as reference for the determination of the color and
  magnitude displacements: MS, horizontal branch, red giant branch
  (RGB) stars, and even variable RR Lyrae \citep[][and references
    therein]{alonco11}. In all cases, the spatial resolution is
  driven by the need for a statistically reliable sample of stars to
build the cell or local CMDs.
To determine the differential extinction map in the direction of
Liller 1 we used the ``star by star'' method, adopting the RGB,
sub-giant branch (SGB) and bright MS stars as reference, as already
tested in a previous work \citep{pallanca6440}.  In addition,
  because of the results discussed in the previous section, we adopted
  R$_V=2.5$.

The first step consists in the definition of the reference mean ridge
line (MRL) of the OP of Liller 1. This has been determined in the
proper motion cleaned, hybrid $(I, I-K)$ CMD, which is significantly
deeper than any other available color-combination (even more than the
$(K, J-K)$ CMD which is likewise deep but with a worst signal to noise
ratio in the $J$ filter at the bottom of the MS; see Figure
\ref{Fig:tripanelNC}), thus allowing us to also use stars along the
SGB and MS phases. In turn, this implies a significant increase of the
spatial resolution of the output reddening map, with respect to the
case where only RGB stars could be used. In order to remove poorly
measured objects that can bias the procedure, we first considered only
stars with high-quality photometry, by adopting 3-$\sigma$ cuts in the
{\it chi} and {\it sharpness} parameter distributions.
The considered sample, comprising the RGB, SGB and the upper MS of the
OP, was divided in magnitude bins of variable size (ranging from 0.25
to 1 mag), in order to best sample the morphological changes of the
sequence in the MS turnoff region. The mean color of each considered
bin was then determined after a $3\sigma$-clipping rejection, in order
to remove any residual interlopers.

Once the MRL is defined, we considered all the stars in the catalog
(i.e., not only the high-quality objects) and for each of them we
selected, from the high-quality pool (and proper motion selected)
sample, the $N_*$ sources spatially lying in the vicinity of the star
under investigation. The method consists in determining the value of
$\delta {\rm E}(B-V)$ required to shift the MRL of the system along
the direction of the reddening vector until it fits the local CMD
defined by these $N_*$ nearby sources. Of course, the larger is the
number of sources ($N_*$), the better defined is the local
CMD. However, to keep the spatial resolution as high as possible, we
constrained $N_*$ below a maximum value and we assumed a limiting
radial distance ($r_{\rm lim}$) within which the nearby sources can be
selected.  We set $N_{*,{\rm max}}=30$ and $r_{\rm lim}= 3\arcsec$.
As soon as one of these two conditions is satisfied, the other one is
no longer considered. Since the stellar density is higher in the
center and decreases with radius, this approach naturally yields an
``adaptive grid'' producing reddening maps with higher (lower) spatial
resolution in the central (external) regions.

As described in F21, the sub-sample with measured PM represents about
$80\%$ of the entire sample. In particular due to dithering patterns
adopted in the GeMS/GSAOI observations (aimed at covering the
inter-chip gaps), a sort of external corona and a central cross of the
FOV was not properly sampled by all the considered NIR exposures to
allow accurate PM measures.  Thus, the procedure aimed at the
construction of the reddening map has been applied twice. The first
time, only stars classified as Liller 1 members from their measured
proper motion were included in the high-quality pool.  The second
time, in order to cover the entire FOV, no proper motion selection was
used.  The differential reddening estimates obtained from these two
approaches are fully compatible (as already found for NGC 6440 in
\citealt{pallanca6440}), demonstrating that the potential presence of
a few field interlopers does not affect the result. Hence, the final
differential reddening value assigned to each star has been obtained,
whenever possible, from the (formally more rigorous) proper motion
based method, and complemented with the value derived neglecting the
proper motion selection in the other cases.

Because of the large central densities, the number of stars within a
circle of $3\arcsec$ radius is typically much larger than 30 in the
inner cluster regions; hence the $N_{*,{\rm max}}$ closest sources are
typically selected within a distance much smaller than $r_{\rm lim}$
from each star under investigation in this region. The opposite is
true in the outskirts, where the driving parameter for the selection
of nearby sources becomes $r_{\rm lim}$. This is clearly visible in
the left panel of Figure \ref{Fig:map}: the distance used for the
selection of the nearby stars defining the local CMD can be as small
as $1\arcsec$ in the central regions, while it increases up to the
maximum value ($r_{\rm lim}=3\arcsec$) close to the boundaries of the
analysed FOV, where only semicircles toward the center can be
drawn. The few spots visibile on the map correspond to local increases
of the selection radius due to the presence of saturated stars. The
discontinuities in the innermost spot and at the outermost edge of the
FOV are due to the transition between reddening estimates that take
into account/neglect the proper motion condition (which is unavailable
in these regions: by keeping larger numbers of stars, this allows us
to use a slightly smaller selection radius).

The comparison between the MRL of the cluster OP and the local CMDs
has been performed in the magnitude range $19.5<I<26$, in order to
discard potentially saturated and too faint (hence not well measured)
objects. We also performed a $3\sigma$ rejection in order to exclude
from the analysis all the stars with a color distance from the MRL
significantly larger than that of the bulk of the selected
objects. This is useful to exclude field interlopers not removed by
the proper motion analysis and/or discard non canonical stars with
intrinsic colors different from those of the cluster main populations
\citep{pallanca10, pallanca13, pallanca14, cadelano15a, cv6624,
  ferraro15, ferraro16b, ferraro18}.

The comparison was performed by shifting the MRL along the reddening
direction (which is fixed by the adopted extinction law) in steps of
$\delta {\rm E}(B-V)$. For each considered step, we calculated the
residual color $\Delta IK$ as:
\begin{equation}\label{Eq:eq1}
  \Delta IK=\sum_{i=1}^{N_*}\left( \left | {IK_{{\rm obs},i} -IK_{{\rm MRL},i}} \right |
  + w_i \cdot \left | {IK_{{\rm obs},i} -IK_{{\rm MRL},i}} \right | \right),
\end{equation}
where $IK_{{\rm obs},i}$ is the $(I-K)$ color observed for each of the
$N_*$ selected objects constituting the local CMD, $IK_{{\rm MRL},i}$
is that of the MRL at the same level of magnitude, and the weight
$w_i$ takes into account both the photometric error on the color
($\sigma$) and the spatial distance ($d$) of the $i^{th}$ source from
the star under investigation, according to the following expression:
\begin{equation}
w_i= \frac{1}{d_i\cdot \sigma_i}
\left[{\sum_{j=1}^{N_*}\left(\frac{1}{d_j\cdot \sigma_j} \right)}
  \right]^{-1}.
\end{equation}
The second term is meant to give more weight to the closest stars,
thus increasing the accuracy of the local reddening estimate, reducing
the impact of using large selection distances where the level of dust
absorption could change.  Note that we cannot reduce equation
(\ref{Eq:eq1}) to the second term only, because a bright object (with
small photometric errors) close to the central star would dominate the
weight and make the reddening estimate essentially equal to the value
needed to move the MRL exactly on that object.  The final differential
reddening value assigned to each star is the value of the step $\delta
{\rm E}(B-V)$ that minimises the normalised residual color ($\Delta
IK/ N_*$).  The spatial distribution of differential extinction values
thus derived for each star constitutes the differential reddening map
shown in the right panel of Figure \ref{Fig:map}.
 
We emphasize that in the adopted procedure, the value of $\delta{\rm E}(B-V)$ assigned to each star is not guided by its position in the CMD with respect to the cluster MRL, but by the position of its $N_*$ neighboring stars (i.e., the $N_*$ stars spatially adjacent on the plane of the sky to the object under analysis). This ensures that we are correcting for a phenomenon (differential reddening) affecting the region of sky where the star is located, not for possible intrinsic properties of the star itself (e.g., its chemical abundance).

\section{DISCUSSION AND CONCLUSIONS}
\label{discuss}
Following the procedure described above, we built the differential
reddening map in the direction of Liller 1, over a total FOV of about
$90 \times 90\arcsec$, and with an angular resolution smaller than
3\arcsec (right panel of Figure \ref{Fig:map}). It appears patchy on
spatial scales of just a few arcsecs, and it presents two main
highly-reddened blobs located on the east and north sides of the
sampled FOV. The less absorbed region (lighter color in the figure) is
located in an almost vertical strip on the west side, flowing into the
north-west and south-east corners. Overall, the map graphically
demonstrates a large amount of differential reddening, with $\delta
{\rm E}(B-V)$ ranging between -0.57 to 0.37.
This corresponds to differential absorptions of  2.16, 1.35, 0.60 and
0.25 mag in the $V$, $I$, $J$ and $K$ filters, respectively.  Note
that the maximum variation of E$(B-V)$ found here ($\sim 0.9$) is
significantly larger than that estimated in \citet{saracino15}, who
quote $\delta E(B-V)=0.34$.   This is however expected both because of
the larger E$(B-V)$ (for a constant percentage of variation, the
larger is the absolute value, the larger is the variation), and
because of the different spatial resolution of the two reddening maps.
In fact, the much larger cells ($18\arcsec\times18\arcsec$) in the map
of \citet{saracino15} have the effect of smoothing local fluctuations,
hence decreasing the amplitude of the E$(B-V)$ variations.

\begin{figure*} 
\begin{center}
\includegraphics[width=190mm]{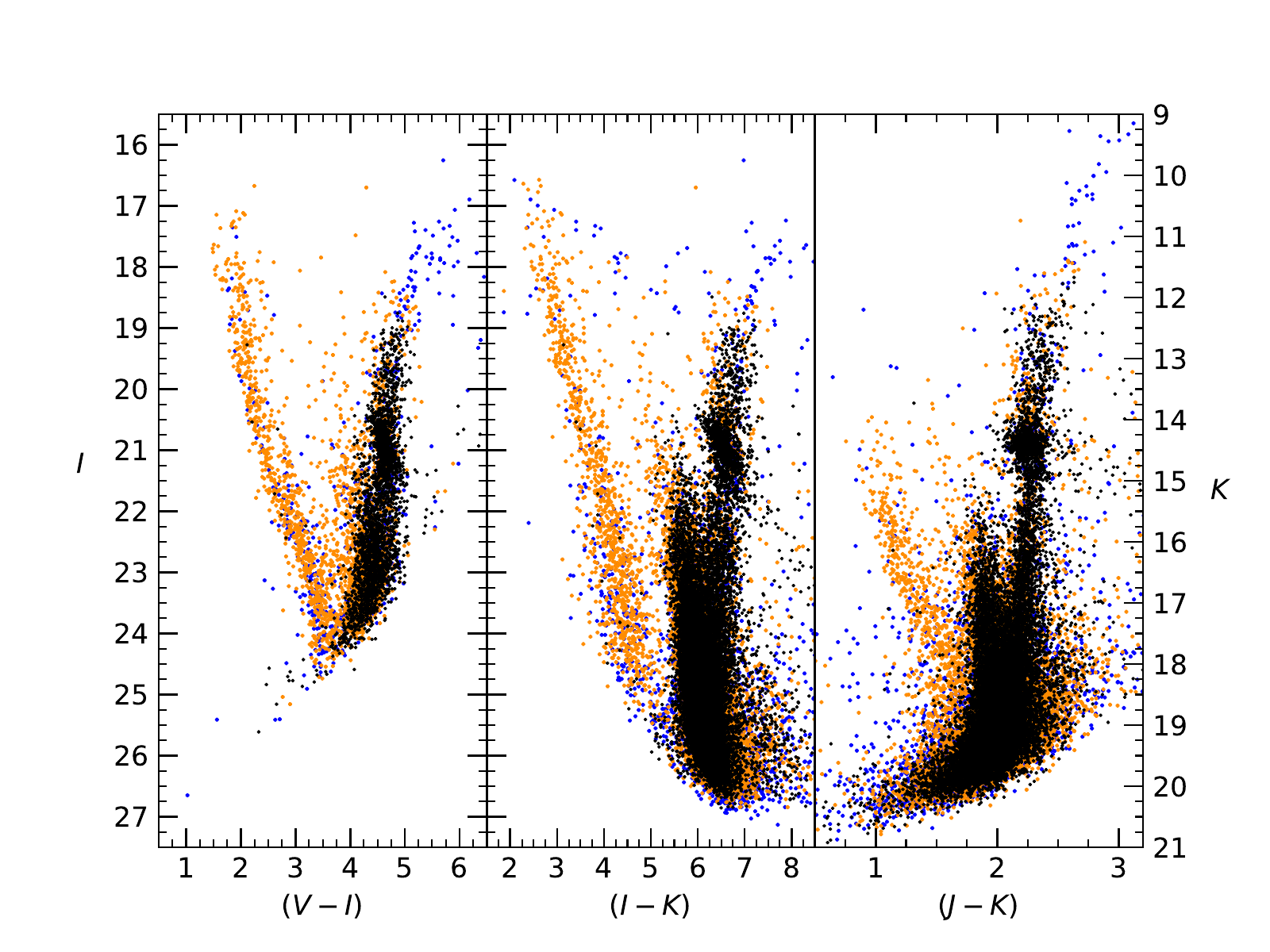}
\caption{CMDs of Liller 1 obtained from the combination of optical HST
  data and GeMS NIR observations  in the common field of view (the green
  area in Figure \ref{Fig:fov}): from left to right, the purely
  optical $(I, V-I)$ CMD, the hybrid $(I, I-K)$ CMD, and the purely NIR
  $(K, J -K)$ diagram.  Black and orange dots mark, respectively, the
  stars considered to be Liller 1 members and Galactic field
  interlopers, according to the proper motion analysis presented in
  Dalessandro E. et al. (2021; in preparation). The blue dots
  correspond to stars for which the proper motion measure is not
  available.  The effect of differential reddening is well evident
  from the elongated morphology of the red clump, especially in the
  optical and hybrid CMDs. 
  }
\label{Fig:tripanelNC}
\end{center}
\end{figure*}

\begin{figure*} 
\begin{center}
\includegraphics[width=190mm]{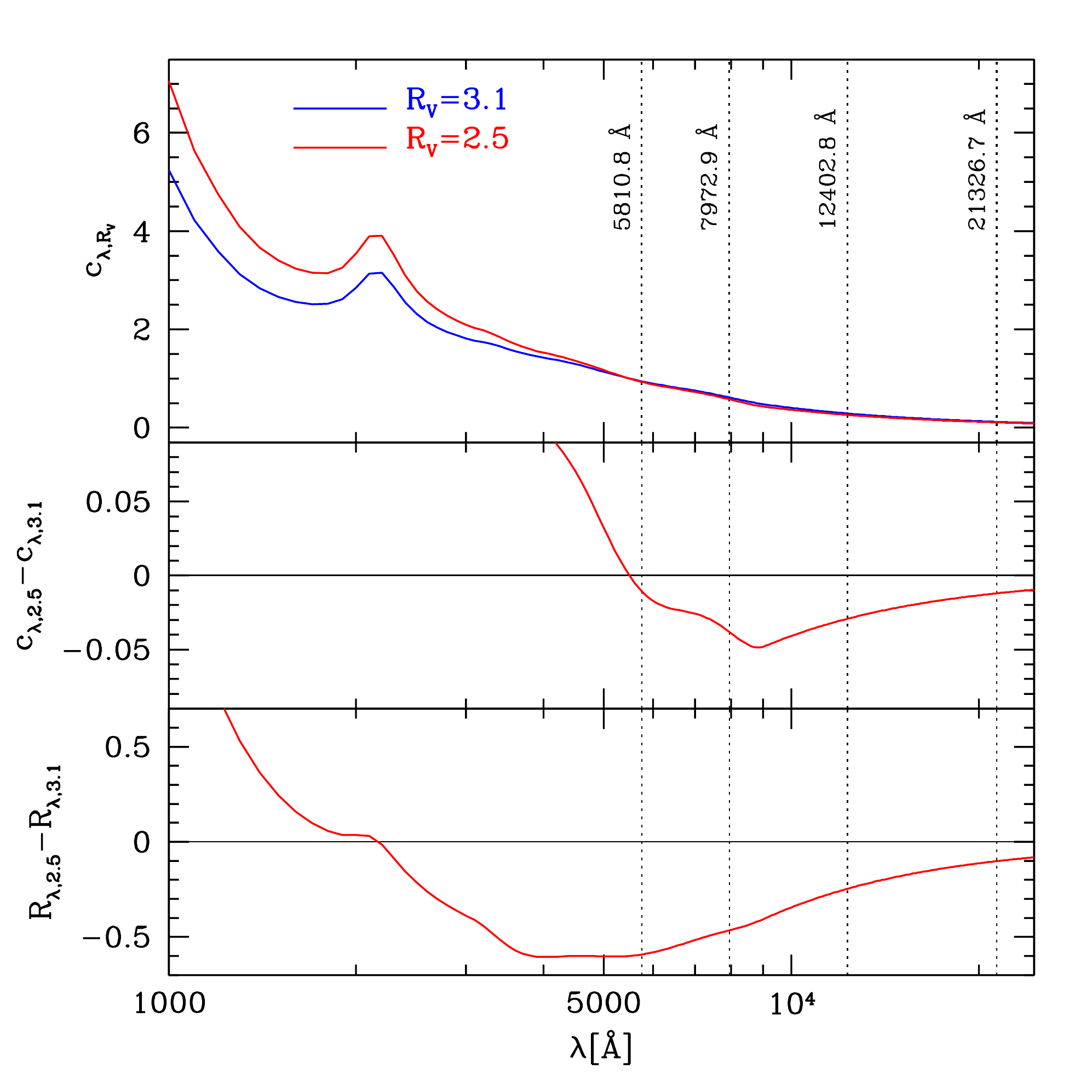}
\caption{{\it Upper panel:} Dependence of the extinction law
  $c_{\lambda, {\rm R}_V}$ (or A$_\lambda/$A$_V$) on wavelength for
  two different assumptions of the R$_V$ coefficient: the standard
  value for diffuse interstellar medium R$_V=3.1$ (blue line), and
  R$_V=2.5$ (red line). The vertical dotted lines are drawn in
  correspondence of the central wavelengths of the four photometric
  filters used in this work (see labels). {\it Middle panel:}
  Difference between $c_{\lambda,2.5}$ and $c_{\lambda,3.1}$ as a
  function of wavelength. {\it Lower panel:} Difference between
  R$_{\lambda,2.5}$ and R$_{\lambda,3.1}$ as a function of wavelength,
  with R$_{\lambda,{\rm R}_V}$ defined as the product between R$_V$
  and $c_{\lambda, {\rm R}_V}$, or the ratio between the absorption
  coefficient A$_\lambda$ and the color excess E$(B-V)$.}
\label{fig_RV}
\end{center}
\end{figure*}

\begin{figure*} 
\begin{center}
\includegraphics[width=190mm]{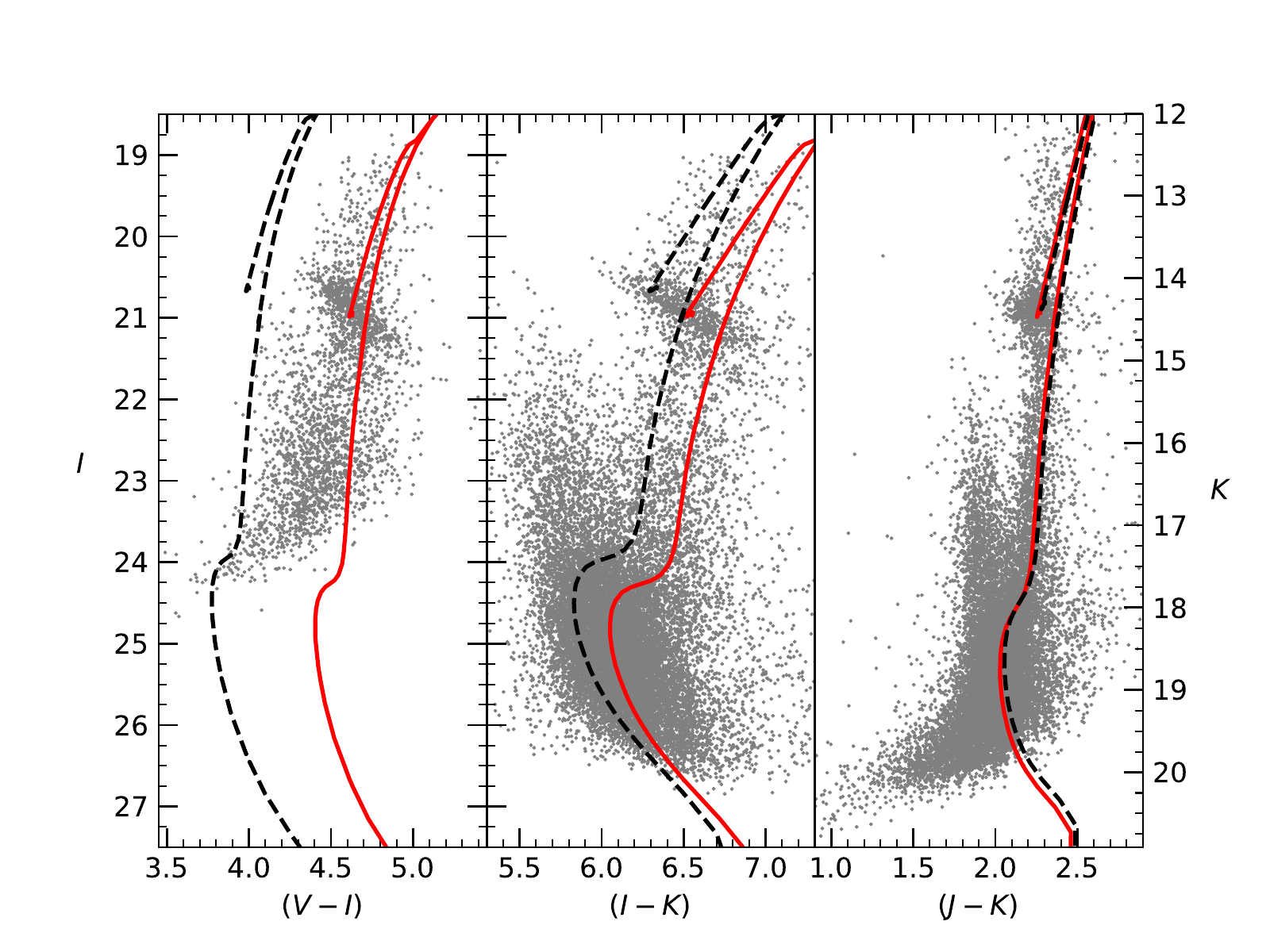}
\caption{ Optical, hybrid and NIR CMDs of Liller 1 obtained from
    well-measured and PM-selected member stars (grey dots), with a 12
    Gyr old, [M/H]$=-0.3$ isochrone superposed. The black dashed line
    marks the position of the isochrone in the three CMDs obtained
    under the assumption of R$_V=3.1$ by requiring that it well
    reproduces the NIR CMD: the corresponding values of mean absolute
    color excess and distance modulus are: E$(B-V)=3.30$ and
    $\mu_0=14.55$ \citep{saracino15}. It is clearly unable to also
    match the hybrid and optical CMDs.  The red solid line marks the
    position of the isochrone under the assumption of R$_V=2.5$ and
    the request to well match the black dashed line in the NIR CMD,
    implying E$(B-V)=4.52$ and $\mu_0=14.65$.  As apparent, it well
    reproduces the observations in all the available color
    combinations.}  
\label{fig_isoCMDobs}
\end{center}
\end{figure*}

\begin{figure*}
\begin{center}
\includegraphics[width=190mm]{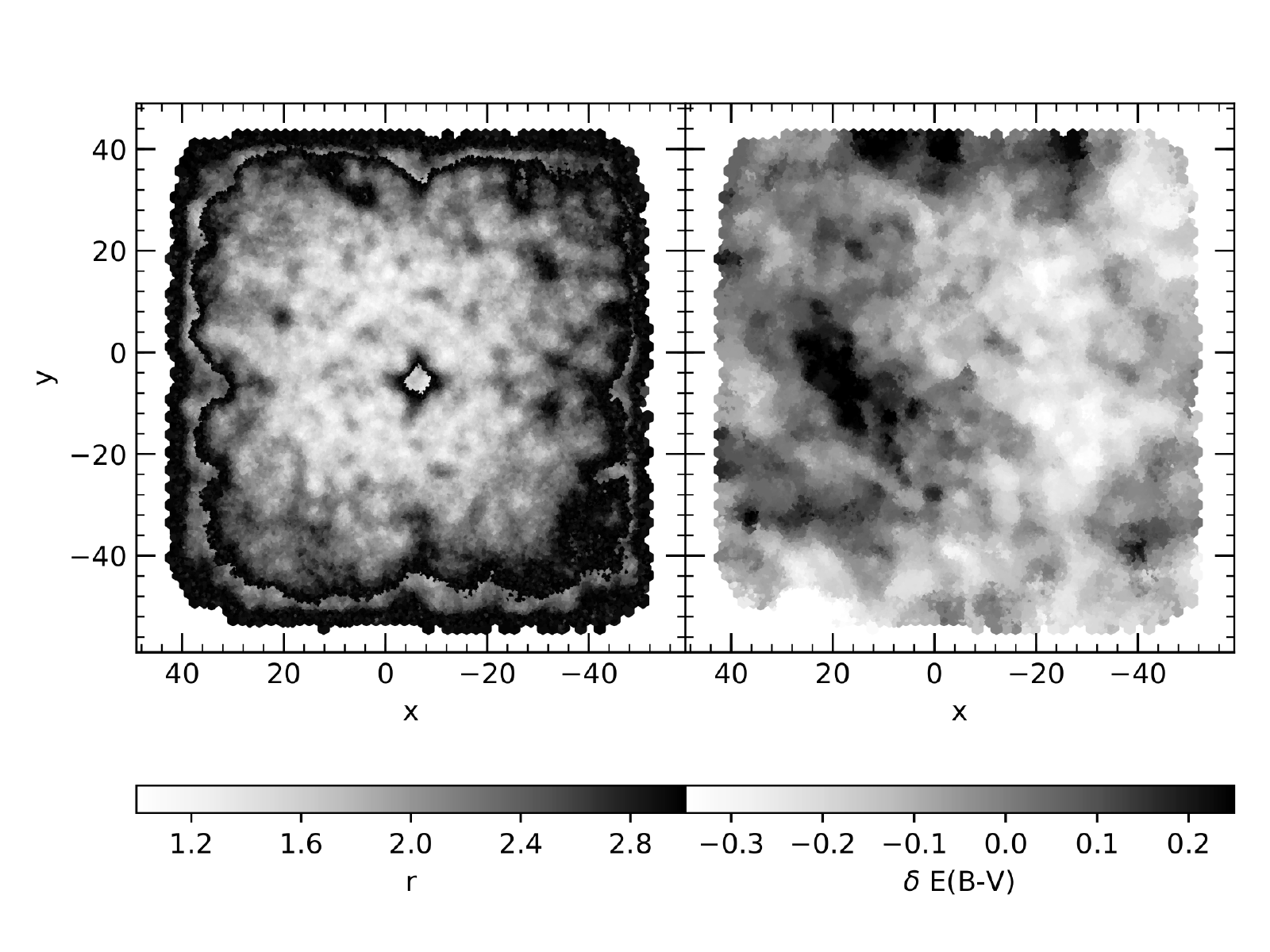}
\caption{{\it Left panel:} Map of the radial distances ($r$, arcsec) used to select
  the samples of nearby sources for the construction of the ``local
  CMD'' of every star in the procedure adpoted to determine the
  differential reddening map. Lighter colors are associated to the
  best reached spatial resolution ($\sim 1 \arcsec$), while  darker
  colors correspond to the regions where $r_{\rm lim}=3\arcsec$ is
  reached.  The discontinuities in the $r$ distribution are due to the
  two different catalogs (with and without the proper motion
  information) used to estimate the differential reddening. See
  Section \ref{diff_redd} for more details.  {\it Right panel:}
  Differential reddening map.
The values of $\delta {\rm E}(B-V)$ are both negative and positive
since they are referred to the MRL of the OP. The associated colorbar
codifies the $\delta {\rm E}(B-V)$ values ranging from less extincted
regions  (lighter colors) to more extincted areas (darker colors).   }
\label{Fig:map}
\end{center}
\end{figure*}

\begin{figure*} 
\begin{center}
\includegraphics[width=190mm]{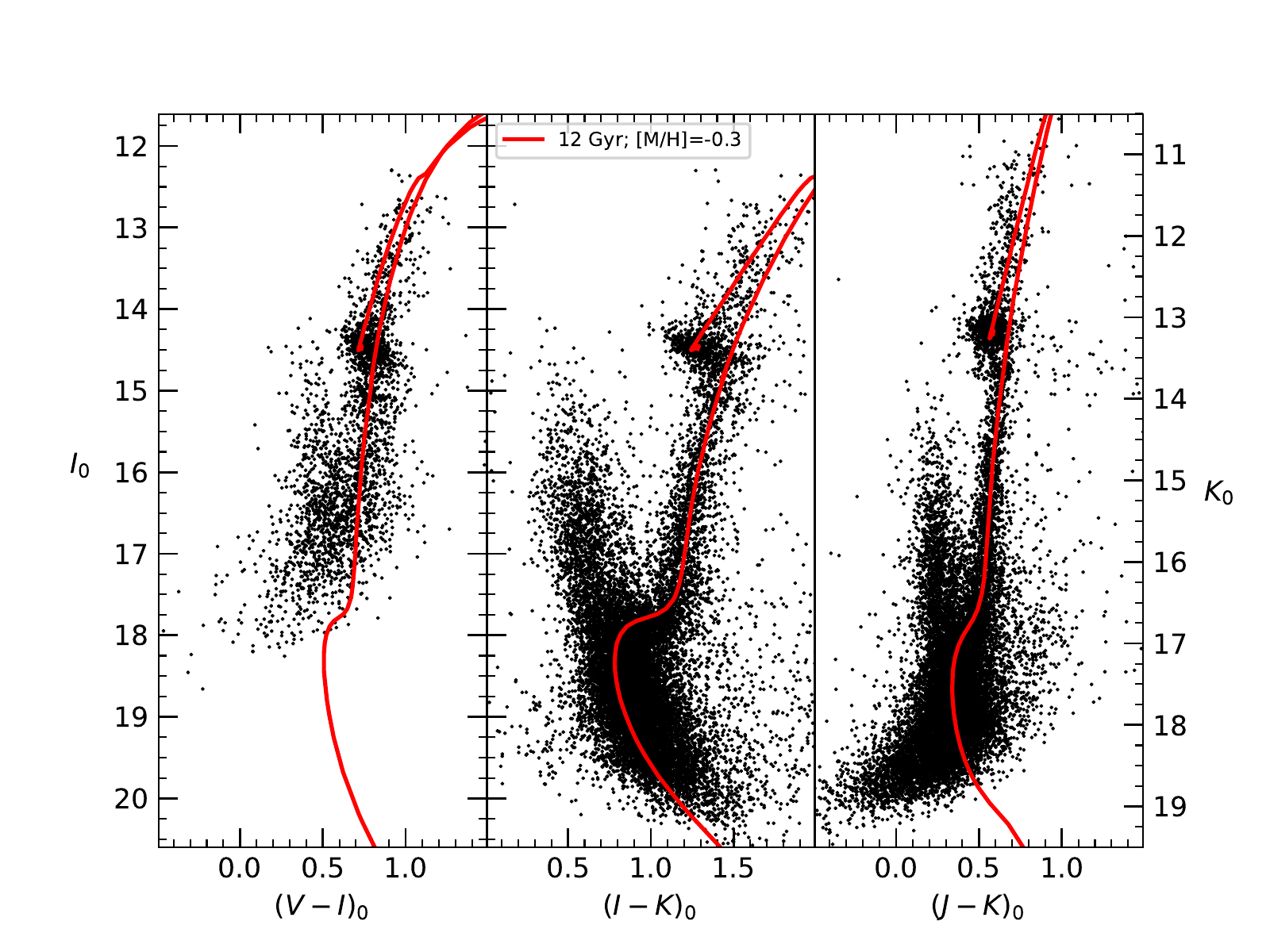}
\caption{ CMDs of of well-measured and PM-selected stars of Liller
    1 after the correction for absolute differential reddening. A 12
    Gyr old isochrone with a metallicity [M/H]$=-0.3$, obtained under
    the assumption of R$_V=2.5$, E$(B-V)=4.52$ and $\mu_0=14.65$ (same
    as the red lines in Figure \ref{fig_isoCMDobs}) is also shown in
    all the panels.}
\label{Fig:tripanelCtot}
\end{center}
\end{figure*}

The CMDs corrected for absolute differential reddening are shown in
Figure \ref{Fig:tripanelCtot} and reveal a clear improvement in their
overall quality with respect to the observed ones (Figure
\ref{Fig:tripanelNC}). Indeed, after correction, the evolutionary
sequences appear sharper (i.e., thinner and better defined) in all the
filter combinations.  In particular, the YP is much more evident and
much better separated in color with respect to the OP (see also
F21). To quantify the effectiveness of the correction, we estimated
the width in the ($I-K$) color at the the level of the faint RGB
($15<K<16$) and of the red-clump ($14<K<14.7$) before and after the
correction: the dispersion in color decreases from $\sim0.2$ mag, down
to $\sim0.1$ mag in both cases.   The position in the reddening
  corrected CMDs of a 12 Gyr old isochrone with [M/H]$=-0.3$, under
  the assumption of R$_V=2.5$, a mean absolute color excess
  E$(B-V)=4.52\pm 0.10$ and a distance modulus $\mu_0=14.65\pm 0.15$
  (see F21) is marked by the red lines (which are the same as in
  Figure \ref{fig_isoCMDobs}).  The very good match of the
  observations in all the three color combinations clearly confirms
  that R$_V=2.5$ is the most appropriate value in the direction of
  Liller 1, while the assumption of the standard value (R$_V=3.1$)
  brings to a severe disagreement in the optical and in the hybrid
  CMDs (analogously to what is shown in Figure \ref{fig_isoCMDobs}).
  The quoted value of $\mu_0$ is consistent with the latest
  determination \citep{saracino15}, while the average reddening
  estimated here is significantly larger than previously found (e.g.,
  3.3 in \citealt{saracino15}).  However, the difference is almost
  entirely ascribable to the different value of R$_V$ adopted here and
  in previous works, and it becomes significantly smaller in terms of
  the absorption coefficients: for instance, here we have A$_K= 2.5
\times 0.107 \times 4.52= 1.21$ and A$_J= 2.5 \times 0.257 \times
4.52= 2.90$, while \citet{saracino15} adopted  A$_K= 3.1 \times 0.113
\times 3.3= 1.16$ and A$_J= 3.1 \times 0.281 \times 3.3= 2.86$, which
are well consistent with our values.

\begin{figure*}
\begin{center}
\includegraphics[width=190mm]{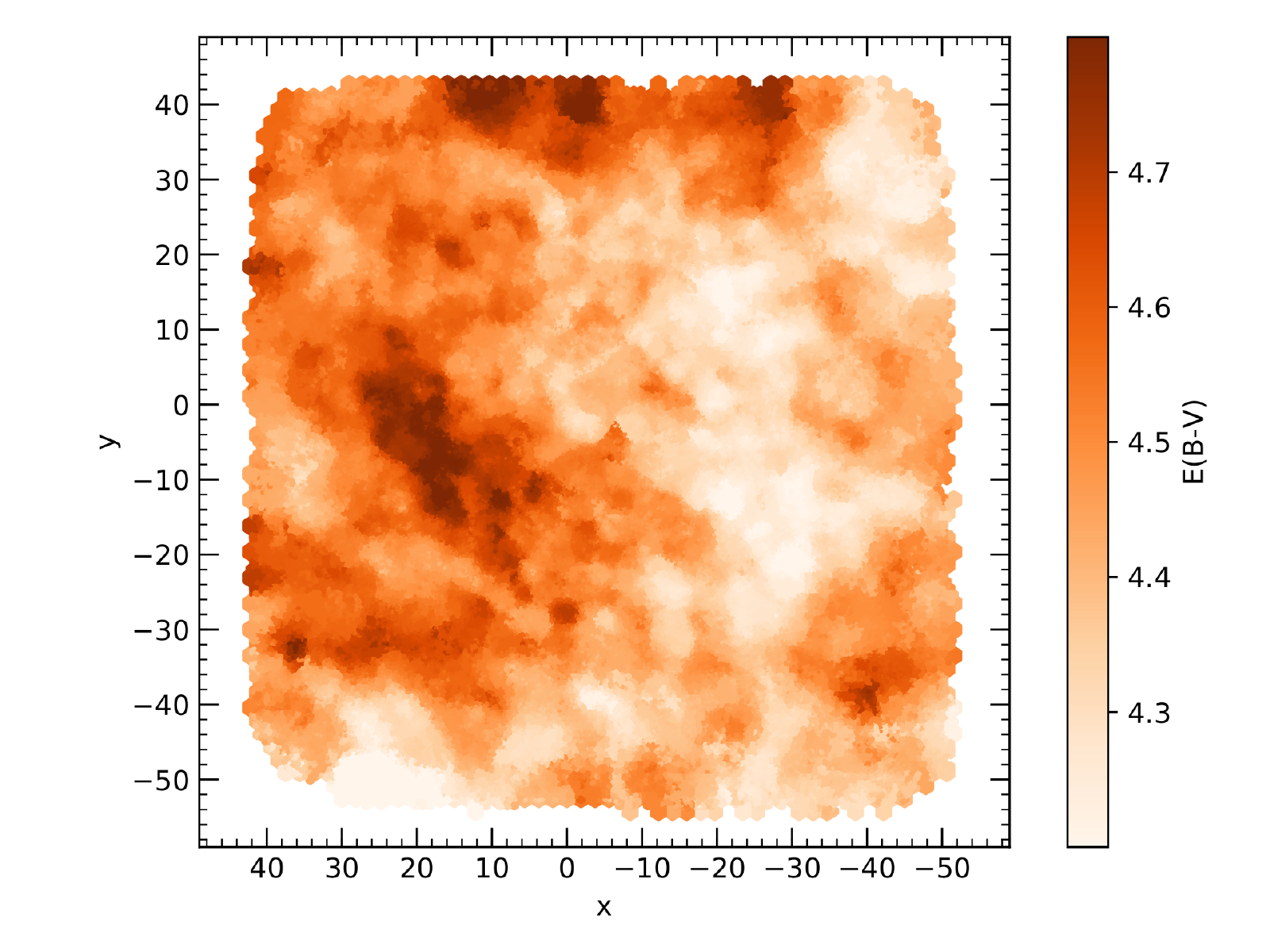}
\caption{Absolute reddening map in the direction of Liller 1. The
  associated colorbar codifies the absolute values of E$(B-V )$, from
  less extincted regions (lighter colors), to more absorbed areas
  (darker colors).}
\label{Fig:map_final}
\end{center}
\end{figure*}

Once the average reddening E$(B-V)$ is known, the absolute reddening
map in the direction of Liller 1 can be constructed. This is shown in
Figure \ref{Fig:map_final}, where different values of the absolute
color excess E$(B-V)$ are coded as in the plotted color-bar. We stress
again that while the absolute values of E$(B-V)$ depend on the adopted
absorption coefficients and extinction law, the main structures
distinguishable in the derived map are largely independent of them.

\begin{figure*} 
\begin{center}
\includegraphics[width=190mm]{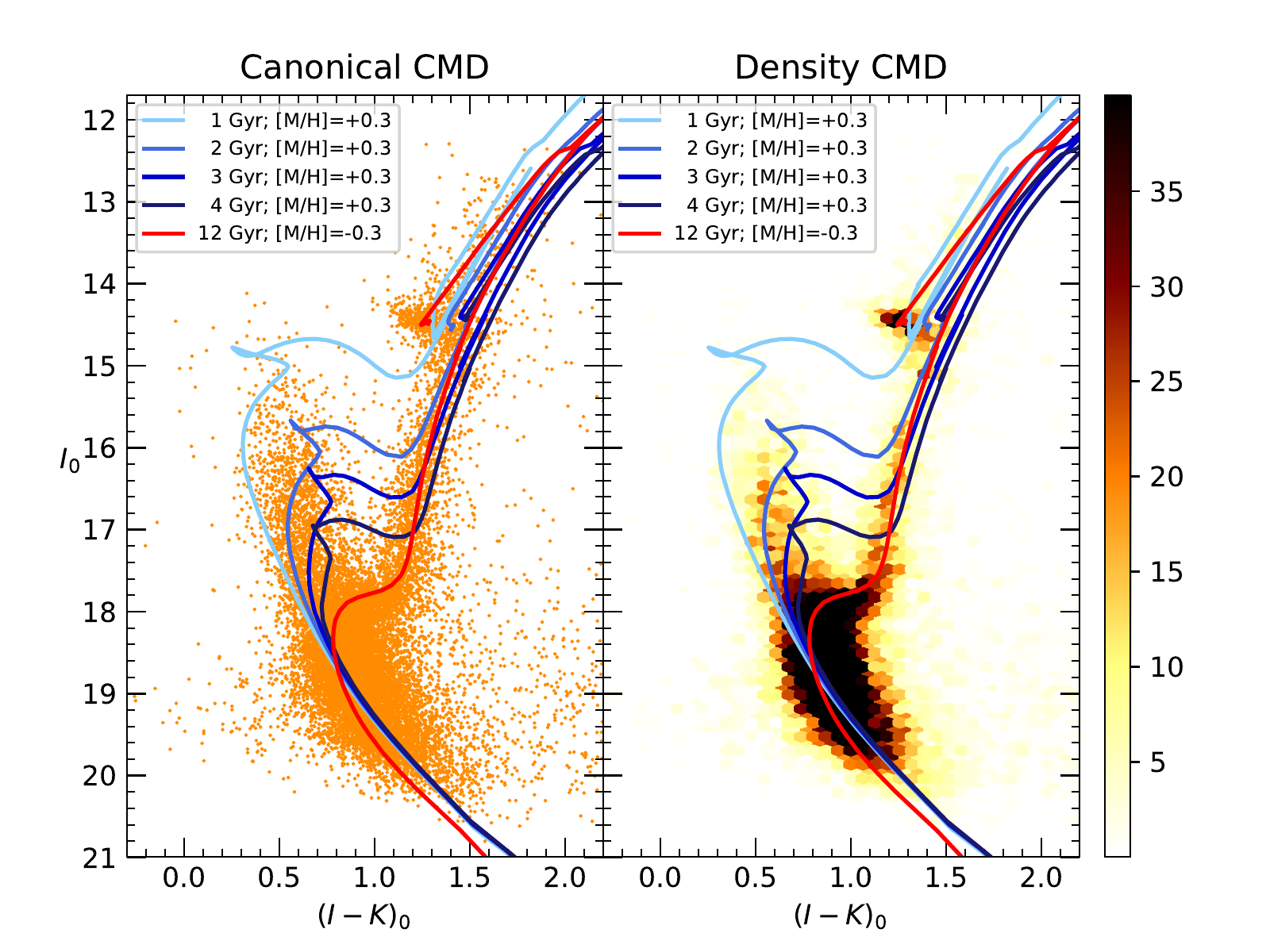}
  \caption{Differentially reddening corrected hybrid CMD with
    isochrones of different ages and metallicities (see labels)
    overplotted.  The left panel shows a canonical CMD (where every
    dot is a resolved star), while the right panel provides a density
    CMD (Hess diagram), where darker colors correspond to larger
    stellar densities in the CMD (see the color-bar on the
    right). Both visualisations highlight the clear presence of a 1-3
    Gyr old stellar component, co-existing with an old (12 Gyr) and
    likely more metal-poor population.}
\label{Fig:cmdISO}
\end{center}
\end{figure*}

 The CMDs shown in Figure \ref{Fig:tripanelCtot} are the cleanest
  and most accurate ever obtained for Liller 1, and they led F21 to
  the discovery of a 1-3 Gyr old stellar population (the YP)
co-existing with the OP in this system. This is illustrated in two
complementary ways in Figure \ref{Fig:cmdISO}.  The left panel shows a
``canonical'' CMD, where each dot corresponds to an individual
star. It provides a clear distinction of the main evolutionary
sequences of the OP, and allows a clean separation of the YP from the
RGB of the OP. The right panel, instead, shows the stellar density
distribution in the CMD (Hess diagram), with the darker colors being
associated to regions of larger numbers of stars per color-magnitude
cell. This way immediately allows the identification of the most
populated sequences (e.g., the MS), making clearer the transition
between the MS and the highly dense SGB of the OP (which are both
typical of very old stellar populations), and the much more prominent
MS of the YP, extending to brighter magnitudes and bluer colors, as
expected for younger stellar populations.  The analysis of this CMD
and the comparison with different sets of isochrones suggest that the
YP is made of stars not only with an age of just 1-3 Gyr, but probably
also with a higher metallicity than the OP (see the isochrones plotted
in Figure \ref{Fig:cmdISO} and the discussion in F21). This finding
unquestionably demonstrates that Liller 1 is not a GC, but a stellar
system with a much more complex star formation history. This is the
second case (after Terzan 5; \citealp{ferraro09,ferraro16a}) of a
GC-like system in the Galactic bulge showing these peculiarities.  F21
demonstrated that the oldest stellar populations observed in Liller 1
and in Terzan 5 are impressively similar, thus indicating that these
systems formed at the same cosmic epoch, from gas clouds with
comparable chemistry\citep{ferraro09, origlia11, origlia13,
  massari14}.  In turn, the striking
similarity between the chemical patterns measured in Terzan 5 and
those observed in the bulge field stars suggests that these systems
possibly are the remnants of massive clumps that contributed to form
the bulge at the epoch of the Galaxy assembling.  The identification
of the BFFs as sites of recent star formation also has a crucial
impact on the presence and the origin of young stars in the Galactic
bulge field.  As shown in this paper the correct evaluation of the
reddening map in the direction of these highly obscured stellar
systems is a crucial requirement to probe their stellar populations
and recontruct their star formation history (Emanuele Dalessandro et
al., in preparation).

\acknowledgments 
 We thank the anonymous referee for the useful comments that improved the paper.
This work is part of the project Cosmic-Lab at the
Physics and Astronomy Department of the Bologna University. The
research was funded by the Italian MIUR throughout PRIN-2017 grant
awarded to the project 2017K7REXT {\it Light-on-Dark} (PI:Ferraro).
We also thank financial support from the Mainstream project SC3K -
Star clusters in the inner 3 kpc (1.05.01.86.21) granted by INAF.  
This research has made use of the SVO Filter Profile Service (http://svo2.cab.inta-csic.es/theory/fps/) supported from the Spanish MINECO through grant AYA2017-84089.
SS gratefully acknowledges financial support from the European Research
Council (ERC-CoG-646928, Multi-Pop).  D.G. gratefully acknowledges
support from the Chilean Centro de Excelencia en Astrof\'isica y
Tecnolog\'ias Afines (CATA) BASAL grant AFB-170002.  D.G. also
acknowledges financial support from the Direcci\'on de Investigaci\'on
y Desarrollo de la Universidad de La Serena through the Programa de
Incentivo a la Investigaci\'on de Acad\'emicos (PIA-DIDULS).

\vspace{5mm}
 
The photometric data that support the plots and other findings of this study are
available from the corresponding author upon reasonable request.
\facilities{HST(ACS-WFC); GeMS/GSAOI}.
\software{\texttt{Matplotlib} \citep{matplotlibref}; 
\texttt{NumPy} \citep{numpyref}; 
\texttt{extinction} \citep{extinction}.}

\end{document}